\documentclass[conference]{IEEEtran}
\IEEEoverridecommandlockouts
\usepackage{cite}
\usepackage{amsmath,amssymb,amsfonts}
\usepackage{soul}
\usepackage{algorithmic}
\usepackage{graphicx}
\usepackage{textcomp}
\usepackage[table]{xcolor}
\usepackage{amsfonts}
\usepackage{booktabs}
\usepackage{float}
\usepackage{graphicx}
\usepackage{grffile}
\usepackage{longtable}
\usepackage{wrapfig}
\usepackage[top=0.75in, bottom=1in, left=0.725in, right=0.725in]{geometry}
\usepackage{gensymb}
\usepackage[utf8]{inputenc}
\usepackage{textgreek}
\usepackage{textcomp}
\usepackage{multicol}
\usepackage{amsmath}
\usepackage{algorithm}
\usepackage{verbatim}
\usepackage{url}
\usepackage{graphicx}
\usepackage{caption}
\usepackage{listings}
\usepackage{array}
\usepackage{subcaption}
\usepackage{booktabs}
\usepackage{amsmath}
\usepackage{enumitem}
\usepackage{dblfloatfix}
\usepackage{balance}
\usepackage{lastpage}
\usepackage{lipsum}
\usepackage{tikz}
\usepackage{array}
\usepackage[utf8]{inputenc}
\usepackage{csquotes}
\usepackage[most]{tcolorbox}

\usepackage{listings}
\newcommand{\cppcode}[1]{\lstinline[language=c++]{#1}}

\newcommand{\blackcirc}{\tikz{\draw[fill=black]  circle(0.1);}}
\newcommand{\whitecirc}{\tikz{\draw[fill=white]  circle(0.1);}}
\newcommand{\halfcirc}{\tikz{\draw[fill=white] (0,0.1) arc[start angle=0, end angle=180, radius=0.1]; \draw (-0.2,0.1) -- (0,0.1); }}
\newcommand{\threequartercirc}{\tikz{\draw[fill=white] (0,0) arc[start angle=0, end angle=270, radius=0.1]; \draw (-0.1,0) -- (0,0);  \draw (-0.1,0) -- (-0.1,-0.1);  }}
\newcommand{\quartercirc}{\tikz{\draw[fill=white] (0,0) arc[start angle=180, end angle=270, radius=0.15]; \draw (0.15,0) -- (0,0); \draw (0.15,0) -- (0.15,-0.15);   }}

\newcommand{\orcid}[1]{\href{https://orcid.org/#1}{\includegraphics[height=10pt]{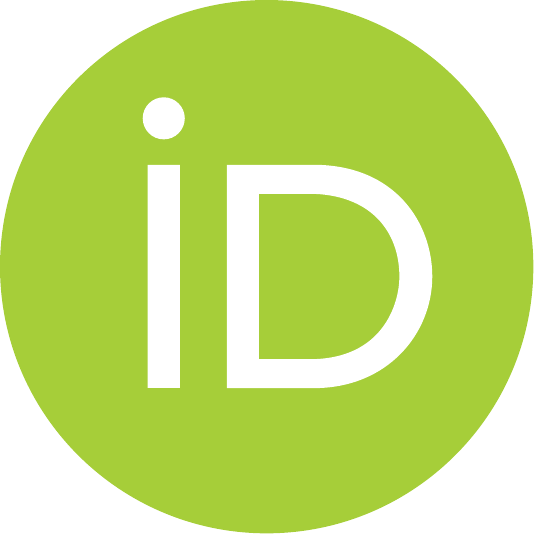}}}

\begin{document}
\title{Can LLMs Find Bugs in Code? An Evaluation from Beginner Errors to Security Vulnerabilities in Python and C\texttt{++}}

\author{
\IEEEauthorblockN{
Akshay Mhatre\IEEEauthorrefmark{1}, Noujoud Nader\IEEEauthorrefmark{2}  , Patrick Diehl\IEEEauthorrefmark{3}\IEEEauthorrefmark{4}, Deepti Gupta\IEEEauthorrefmark{1}
}
\IEEEauthorblockA{
\IEEEauthorrefmark{1}Dept. of Computer Information Systems, Texas A\&M University - Central Texas, TX, 76549 USA.}
\IEEEauthorblockA{
\IEEEauthorrefmark{2}LSU Center for Computation \& Technology, Louisiana State University,
Baton Rouge, LA, 70803 USA.}
\IEEEauthorblockA{\IEEEauthorrefmark{3} Department of Physics and Astronomy, Louisiana State University, Baton Rouge, LA, 70803 USA.}
\IEEEauthorblockA{\IEEEauthorrefmark{4} Applied Computer Science (CCS-7), Los Alamos National Laboratory, Los Alamos, NM 87545 USA.
}
E-Mail: am271@my.tamuct.edu, nnader@lsu.edu, diehlpk@lanl.gov, and d.gupta@tamuct.edu
}

\maketitle
\begin{abstract}
Large Language Models (LLMs) such as ChatGPT-4, Claude\ 3, and LLaMA\ 4 are increasingly embedded in software/application development, supporting tasks from code generation to debugging. Yet, their real-world effectiveness in detecting diverse software bugs, particularly complex, security-relevant vulnerabilities, remains underexplored. This study presents a systematic, empirical evaluation of these three leading LLMs using a benchmark of foundational programming errors, classic security flaws, and advanced, production-grade bugs in C++ and Python. The dataset integrates real code from SEED Labs, OpenSSL (via the Suresoft GLaDOS database), and PyBugHive, validated through local compilation and testing pipelines. A novel multi-stage, context-aware prompting protocol simulates realistic debugging scenarios, while a graded rubric measures detection accuracy, reasoning depth, and remediation quality. Our results show that all models excel at identifying syntactic and semantic issues in well-scoped code, making them promising for educational use and as first-pass reviewers in automated code auditing. Performance diminishes in scenarios involving complex security vulnerabilities and large-scale production code, with ChatGPT-4 and Claude\ 3 generally providing more nuanced contextual analyses than LLaMA\ 4. This highlights both the promise and the present constraints of LLMs in serving as reliable code analysis tools.
\end{abstract}


\section{Introduction}
% LLMS
%Large Language Models (LLMs) are a class of Machine Learning (ML) models designed for natural language processing tasks. Notable examples include ChatGPT, LLaMA, GitHub Copilot, Claude, and BERT, which have gained widespread recognition for their ability to perform effectively in a variety of linguistic and reasoning tasks.

%
%Among their many capabilities, LLMs excel at translating natural language instructions into executable code, supporting code comprehension, debugging, and answering code-related queries. These features have become valuable tools for developers, helping to streamline workflows, accelerate development, and improve overall productivity.

%Bug detection

%models and PLS

%s

Large Language Models (LLMs) are a class of Machine Learning (ML) models that have revolutionized natural language understanding and generation~\cite{chang2024survey}. Models such as ChatGPT~\cite{achiam2023gpt}, Claude~\cite{claudeClaude}, GitHub Copilot~\cite{dakhel2023github}, LLaMA~\cite{touvron2023llama}, and BERT~\cite{koroteev2021bert} have become dominant tools in software development scenarios, where these tools are used for tasks ranging from translating natural language instructions into code to performing debugging and answering code-related queries~\cite{diehl2024evaluating,nader2025llmhpcbenchmarkingdeepseeks,diehl2025llm}. Their utility has exploded rapidly on various platforms such as Integrated Development Environments (IDE), educational platforms, and enterprise DevOps pipelines, accelerating programming workflows and lowering barriers to entry for novice developers~\cite{liu2024empirical}. However, the usage of these models and the accuracy of LLMs' potential, especially when applied to diverse, realistic software bugs, remain underexplored. This gap is critical, as undetected bugs can lead to functional failures, security breaches, and substantial maintenance costs in modern software systems.

This study addresses this critical question by conducting a systematic empirical evaluation of three cutting-edge LLMs, namely ChatGPT-4, Claude 3 (Sonnet 3.7), and LLaMA 4 (maverick), on their ability to detect and explain a broad spectrum of software bugs. Our investigation spans foundational programming mistakes, classic security vulnerabilities, and advanced, real-world bugs drawn from widely-used open-source systems. The code samples include C++ snippets from SEED Labs and OpenSSL (via the Suresoft GLaDOS bug database), as well as Python bugs from the PyBugHive repository~\cite{antal2024pybughive}, covering libraries like NumPy and Pandas. Unlike prior studies that often rely on synthetic or narrowly-scoped examples, our dataset integrates checking codes and validating bugs. Moreover, our methodology includes rigorous local validation of each bug, multi-stage prompting to simulate real-world debugging, and a graded detection rubric to measure not just correctness, but also depth of reasoning and remediation quality. This multifaceted evaluation allows us to expose both the strengths and limitations of current LLMs when operating as static and dynamic code analyzers under varied code complexity levels and programming paradigms.

The novelty of this research lies in its context-aware framework for assessing LLM performance across multiple layers of code reasoning, from shallow syntax-based flaws to deeper semantic, contextual, and security-related vulnerabilities. This work is motivated by the increasing dependence on LLMs in automated code review pipelines, educational tools for programming instruction, and even security auditing processes.

The main contributions of this paper are as follows:
\begin{itemize}
\item Developed a comprehensive benchmark composed of foundational, security, and advanced real-world bugs across C\texttt{++} and Python, validated through local compilation and testing pipelines.
\item Designed and implemented a novel, multi-stage, context-aware prompting protocol to mimic realistic developer interaction and debugging workflows.
\item Conducted a comparative, fine-grained analysis of three leading LLMs (ChatGPT-4, Claude 3, and LLaMA 4), assessing detection accuracy, reasoning depth, and suggestion quality using a standardized rubric.
\end{itemize}

The remainder of this paper is organized as follows. Section~\ref{related} presents the literature review on vulnerability detection using LLMs. We present the methodology of this work in Section~\ref{methodology}. Various types of bugs and their detection are discussed in Section~\ref{qual}. Advanced work on bug detection of C\texttt{++} and Python is presented in Section~\ref{advanced1} and Section~\ref{advanced}, respectively. Discussion is presented in Section~\ref{Discuss}. Conclusion and future work are discussed in Section~\ref{conclusion}.
\section{Related Work}
\label{related}
%With respect to vulnerability detection, the following work has been done: CodeQwen1.5, DeepSeek-Coder, CodeGemma, Starcoder-2, and CodeLlama were explored to find vulnerabilities in Python, Java, and Javascript~\cite{zhang2025benchmarking}; \textit{LProtector} used ChatGPT to analyze C\texttt{++} code~\cite{sheng2024lprotector}; and \textit{FuncVul} was used to analyze C/C\texttt{++}~\cite{halder2025funcvul}. Vulnerable code found on \textit{Stack Overflow} was injected into Claude\ 3, GPT-4, and Llama\ 3 to determine if the model detects whether the code is secure or not~\cite{sajadi2025llms}. Another topic of interest is Automated Program Repair (APR) with LLMs. However, the automation is not part of this work yet. \textit{RepairAgent} uses autonomous agents based on a large language model (LLM) to find bugs and fix them~\cite{bouzenia2024repairagent}. \textit{SRepair} fixes multi-function bugs and demonstrates cost efficiency with US\$$0.029$/Fixed Bug~\cite{xiang2024far}. A Systematic Literature Review of LLMs in Code Security is provided in~\cite{basic2024large}. Automated code repair is surveyed here~\cite{zhang2024systematic}.

With respect to vulnerability detection, a substantial body of recent work has focused on evaluating code-oriented LLMs across multiple programming languages and vulnerability types. Several studies systematically benchmark models such as CodeQwen1.5, DeepSeek-Coder, CodeGemma, Starcoder-2, and CodeLlama on Python, Java, and JavaScript code to assess their ability to identify security weaknesses~\cite{zhang2025benchmarking}. These efforts demonstrate that specialization and scale both influence detection performance, with models showing varying sensitivity to language-specific constructs and common vulnerability patterns. However, the benchmarking setup typically emphasizes static snippets and predefined vulnerability categories, which limits insight into how these models generalize to more complex or less structured real-world codebases. Beyond general benchmarking, other work explores targeted applications of LLMs for vulnerability analysis in lower-level languages such as C and C++. For example, LProtector leverages ChatGPT to analyze C++ programs, aiming to identify security issues through natural-language-driven reasoning about program structure and logic~\cite{sheng2024lprotector}. Similarly, FuncVul focuses on function-level analysis of C/C++ code, highlighting the importance of localized semantic understanding when detecting vulnerabilities in systems programming contexts~\cite{halder2025funcvul}. These approaches show that LLMs can reason about memory management, pointer usage, and control flow to some extent, but they also reveal challenges related to incomplete context, false positives, and inconsistent reasoning across different code samples. Several recent studies~\cite{diehl2025llm, karim2025securing, kavuri2025securefed, nair2025androids} in the literature have explored this problem from different perspectives, highlighting its growing research interest.

Another complementary line of work investigates how LLMs assess the security of code sourced from community platforms. Vulnerable code snippets collected from Stack Overflow have been injected into models such as Claude 3, GPT-4, and Llama 3 to evaluate whether the models can correctly judge whether the code is secure or insecure~\cite{sajadi2025llms}. The results suggest that while LLMs can often flag obvious vulnerabilities, they may struggle with subtle or context-dependent flaws and sometimes provide overly confident but incorrect assessments. This variability underscores the need for deeper analysis of model reasoning processes and highlights the risks of deploying LLMs as authoritative security judges without additional safeguards. Closely related to vulnerability detection is the growing interest in Automated Program Repair (APR) using LLMs. Systems such as RepairAgent employ autonomous LLM-based agents to locate bugs and generate corresponding fixes, demonstrating that LLMs can move beyond detection toward actionable remediation~\cite{bouzenia2024repairagent}. SRepair further shows that LLM-driven repair can address multi-function bugs efficiently, reporting notably low cost per fixed bug~\cite{xiang2024far}. Despite these advances, most APR approaches assume that the underlying bug or vulnerability has already been correctly identified, and they often prioritize functional correctness over security guarantees. As a result, full automation of secure program repair remains an open challenge and is not the focus of this work.

Finally, broader perspectives on this research area are provided through survey and review studies. A systematic literature review of LLMs in code security consolidates findings across vulnerability detection, analysis, and mitigation, identifying common strengths, limitations, and open problems~\cite{basic2024large}. In parallel, automated code repair techniques, including those based on LLMs, are comprehensively surveyed to contextualize recent progress within the wider software engineering landscape~\cite{zhang2024systematic}. Together, these works motivate the present research by revealing a clear need for more robust, explainable, and practically grounded approaches to LLM-based vulnerability detection, forming the foundation for future advances in secure automated code analysis.
\section{Methodology}
\label{methodology}

This research examines the capacity of LLMs to detect software bugs, ranging from foundational programming errors to advanced real-world issues in widely-used open-source systems. We evaluate three cutting-edge LLMs---ChatGPT-4, Claude\ 3, and LLaMA\ 4---against a diverse set of C\texttt{++} and Python code snippets, using a consistent evaluation protocol to assess diagnostic accuracy and fix-suggestion performance.

\textit{Dataset and Bug Categories}: The dataset is stratified into three main categories:

\begin{itemize}
    \item \textbf{Beginner Bugs (C\texttt{++} and Python)}: These include fundamental issues commonly encountered in early computer science coursework, such as uninitialized variables, incorrect parameter passing, memory mismanagement, logical oversights, and pointer misuse. These examples are reflective of what students encounter in introductory programming courses at universities and colleges and were selected to assess LLMs’ ability to detect foundational errors with minimal context.

    \item \textbf{Security Vulnerabilities (C\texttt{++})}: This category includes snippets exhibiting classic security flaws such as race conditions, format string vulnerabilities, buffer overflows, unsafe memory access, and privilege escalation vectors. These examples were primarily drawn from SEED Labs, an academic platform for cybersecurity education.

    \item \textbf{Advanced Real-World Bugs}:
    \begin{itemize}
        \item \textit{C/C\texttt{++} Bugs}: Extracted from real issues documented in the OpenSSL project and archived in the Suresoft GLaDOS bug database\footnote{\url{https://github.com/Suresoft-GLaDOS}}. These bugs include memory safety violations, cryptographic misconfigurations, and type inference issues.
        \item \textit{Python Bugs}: Sourced from PyBugHive\footnote{\url{https://pybughive.github.io/}}, a database of manually validated bugs in mainstream Python libraries, particularly NumPy and Pandas. These include internal state inconsistencies, API misuse, data structure conflicts, and edge-case behaviors relevant to scientific computing workflows.
    \end{itemize}
\end{itemize}
\vspace{.1in}

\textit{Validation and Local Execution}
All easy bug snippets were manually compiled and executed on local machines using C\texttt{++} compilers (GCC 7.5) and Python 3.7 interpreters to validate their runtime behavior and confirm the presence of the expected compile-time, runtime, or logical errors. For advanced bugs, the corresponding open-source repositories were cloned, and the bugs were reproduced using the pre-patch versions, as verified through commit history and issue discussions. Associated unit tests and test cases were run locally to confirm that the buggy behavior was both reproducible and verifiable. The buggy version of the code along with minimal contextual dependencies was then provided to the LLMs as input.

\vspace{.1in}
\textit{Prompting and Interaction Protocol}
For each snippet, a fresh chat session was initiated to prevent context carryover or bias from previous prompts. This was especially important to avoid situations where the model would focus solely on one type of bug (\emph{e.g.}, memory leaks) even when the actual issue was logical or structural.

For easy bugs, the following standardized prompt was used:

This minimal prompting approach assessed the LLMs’ baseline ability to reason about standalone code.

\begin{displayquote}
\begin{tcolorbox}[colback=gray!15]
Are there any compile-time, runtime, logical errors, or vulnerabilities in the following code? List them with the lines where they occur.
\end{tcolorbox}
\end{displayquote}

For advanced bugs, the prompt was extended with relevant auxiliary context, such as reduced versions of dependent files or known interfaces. 

If the LLM's first response listed a candidate issue or line, the next prompt would explicitly reference that function or line to probe for a deeper analysis or partial identification. When the LLM failed to surface meaningful results, we further introduced unit test descriptions or inputs, simulating how developers would triangulate buggy behavior using the test suite.

This multi-stage, context-aware prompting was crucial for assessing whether models could go beyond surface-level static checks and demonstrate adaptive reasoning under more realistic, project-scale scenarios.

\begin{displayquote}
\begin{tcolorbox}[
    colback=gray!15, 
    boxrule=0.4pt, 
    arc=3pt, 
    left=6pt, 
    right=6pt, 
    top=4pt, 
    bottom=6pt,
    before skip=6pt,
    after skip=6pt,
    enhanced jigsaw
]

\label{A3 - Python Advance bug} 

Identify all internal logical errors, edge cases, and potential defects within the \textbf{\texttt{PyArray\_FillWithScalar}} function, as well as those that may arise from its usage.

\vspace{0.75em}
\rule{\linewidth}{0.2pt}
\vspace{0.4em}

Based on the following test case, evaluate where \textbf{\texttt{PyArray\_FillWithScalar}} could fail and cause an issue.

\vspace{0.5em}
\begin{lstlisting}[language=Python,
basicstyle=\ttfamily\scriptsize,
]
def test_fill_readonly(self):
    a = np.zeros(11)
    a.setflags(write=False)
    with pytest.raises(
        ValueError, 
        match=".*read-only"
    ):
    a.fill(0)
\end{lstlisting}
\end{tcolorbox}
\end{displayquote}
\vspace{0.25cm}

\subsection*{Evaluation Criteria}

Detection performance was rated using a graded visual scheme:
\begin{itemize}
    \item \whitecirc \quad \textbf{Empty Circle} — \textit{No detection of the bug}
    \item \quartercirc \quad \textbf{Quarter Circle} — \textit{Minimal or speculative hint}
    \item \halfcirc \quad \textbf{Half Circle} — \textit{Partial detection}
    \item \threequartercirc \quad \textbf{Three-Quarter Circle} — \textit{Near-complete detection}
    \item \blackcirc \quad \textbf{Full Circle} — \textit{Complete detection of the bug}
\end{itemize}

\section{Quality of Bug Detection}
\label{qual}
Table~\ref{tab:compiler} lists the compilers and libraries used for the evaluation. We used the standard GCC and Python versions.
%installed by Linux. 

\begin{table}[b]
    \centering
    \begin{tabular}{cc|cc|cc|cc}\toprule
     GCC &  7.5 & Python & 3.7 & numpy & 1.23.4 & pandas & 0.19.2 \\\bottomrule
    \end{tabular}
    \caption{Compiler and software versions used in the study.}
    \label{tab:compiler}
\end{table}

\subsection{Beginners C\texttt{++} Bugs}
\label{qual:easy:cpp}

\begin{table*}[htbp]
\centering
\begin{tabular}{l|ccc}
\toprule
Language C\texttt{++}  & \multicolumn{3}{c}{LLM} \\
\midrule
Bug & GPT & Llama & Claude \\
\midrule
\textbf{E1}: Memory allocation error (\cppcode{std::bad_alloc}) on oversized int & \blackcirc & \blackcirc & \blackcirc \\
\rowcolor{lightgray} \textbf{E2}:  Integer Underflow and Truncation & \blackcirc & \blackcirc & \blackcirc \\
\textbf{E3}:  Pass-by-value prevents actual update & \blackcirc & \blackcirc & \blackcirc \\
\rowcolor{lightgray} \textbf{E4}:  Memory freed twice & \blackcirc & \blackcirc & \blackcirc \\
\textbf{E5}:  Accessing deallocated heap via pointer & \blackcirc & \blackcirc & \blackcirc \\
\rowcolor{lightgray} \textbf{E6}:  \cppcode{std::move} leaves vector in unknown state & \blackcirc & \blackcirc & \blackcirc \\
\textbf{E7}:  Buffer overflow risk (no bounds check) & \blackcirc & \blackcirc & \blackcirc \\
\bottomrule
\end{tabular}
\caption{Evaluation of bugs commonly done by C\texttt{++} novices on the bachelor level.}
\label{tab:evaluation:easy:cpp}
\end{table*}

This section evaluates the ability of LLMs to identify basic C\texttt{++} programming bugs that are representative of issues commonly introduced by novice programmers during early coursework. These bugs were selected to reflect pedagogical examples typically used in \textit{introductory computer science curricula at universities and colleges}, where students learn fundamental programming constructs such as pointer management, reference passing, and basic memory operations. These examples form a baseline for assessing LLMs’ capability to statically reason about localized, self-contained code snippets.

We examined seven such C\texttt{++} snippets (denoted as \textbf{E1} to \textbf{E7}) and submitted them to ChatGPT-4, Claude\ 3, and LLaMA\ 4 under consistent prompt conditions. Each model was asked to identify any compile-time, runtime, logical, or vulnerability-related bugs in isolation and, if possible, suggest a corrective action. All code snippets were additionally compiled and executed locally to \textit{validate their error conditions} and match LLM responses with observed behavior.

\paragraph{ \textbf{E1:} Unbounded Allocation via \cppcode{std::rand()}} This snippet triggers a potential \cppcode{std::bad_alloc} error by using an unbounded call to \cppcode{std::rand()} to reserve a vector size, possibly exceeding available memory. All LLMs correctly identified this risk and recommended bounding the random value using modulo or conditional checks to limit memory allocation, a clear affirmation of their understanding of unsafe dynamic allocation patterns.

\paragraph{ \textbf{E2:} Pass-by-Value Semantics and Side Effects} In this snippet, a template function intended to square a value is passed by value, leading to no effect on the original variable. All models accurately flagged the semantic bug—namely, that pass-by-value prevents mutation of the calling context—and recommended using references to preserve side effects.

\paragraph{ \textbf{E3:} Double Deletion Across Scopes} This example involves deleting the same dynamically allocated memory in both a helper function and in \cppcode{main()}, resulting in undefined behavior due to double deletion. All three LLMs correctly identified the ownership confusion and suggested smart pointer usage, such as \cppcode{std::unique_ptr}, to resolve the issue.

\paragraph{ \textbf{E4:} Use-After-Free Before Function Call} In this case, memory is deallocated before being passed to a function that attempts to iterate over the now-invalid pointer. All models correctly flagged this use-after-free bug, suggesting the pointer be freed after the loop completes.

\paragraph{ \textbf{E5:} Misuse of \cppcode{std::move} and Vector Invalidity} This snippet misuses \cppcode{std::move()} to transfer ownership of a \cppcode{std::vector} into a function, after which its size is queried. The models identified that the moved-from vector's state is unspecified post-transfer and suggested redesigning the call or avoiding post-move usage.

\paragraph{ \textbf{E6:} Null Pointer Dereference} This example attempts to dereference an uninitialized pointer (\\cppcode{nullptr}) in a loop, leading to undefined behavior. All LLMs identified the segmentation fault risk and noted the pointer lacked memory allocation before accessing an example of proper detection of null-pointer safety.

\paragraph{ \textbf{E7:} Buffer Overflow via Unbounded String Copy} A classic buffer overflow scenario, this snippet copies input from \cppcode{argv[1]} into a fixed-size buffer without bounds checking. Claude and LLaMA correctly identified the overflow risk. ChatGPT flagged the problematic line but gave only partial elaboration.

In summary, for these foundational bugs, all LLMs demonstrated strong diagnostic capabilities, routinely identifying the nature, location, and remedy of errors. Their recommendations aligned with \textit{modern C\texttt{++} best practices}, showing fluency in using smart pointers and proper memory ownership semantics. These findings suggest that LLMs are well-suited for beginner-level static analysis tasks and could potentially augment educational tooling in programming pedagogy. Table~\ref{tab:evaluation:easy:cpp} summarizes the results.

\begin{table*}[htbp]
\centering
\begin{tabular}{p{7cm}|ccc}
\toprule
Description & GPT & Llama & Claude \\
\midrule
\rowcolor{lightgray} Snippet number: Description of the issue & GPT & Llama & Claude \\
\textbf{S1}: Format string vulnerability, Buffer Overflow Vulnerability and leaking memory addresses&  \blackcirc & \threequartercirc &  \blackcirc \\
\rowcolor{lightgray} \textbf{S2}: Race-condition vulnerability& \threequartercirc & \threequartercirc & \threequartercirc \\
\textbf{S3}: Leaked file descriptor, weak privilege dropping, unsafe spawning of a shell, and passing a null environment to execve&  \blackcirc & \threequartercirc &  \blackcirc \\
\rowcolor{lightgray} \textbf{S4}: Zip bombs attack, disk exhaustion& \threequartercirc & \threequartercirc &  \blackcirc \\
\textbf{S5}: Out-of-bound and memory leak possibility&  \blackcirc &  \blackcirc &  \blackcirc \\
\rowcolor{lightgray} \textbf{S6}: Temp file creation using temp() and mstemp() functions&  \blackcirc &  \blackcirc &  \blackcirc \\
\midrule
\end{tabular}
\caption{C\texttt{++} Security Bug Evaluation}
\label{tab:cpp:security:bug:evaluation}
\end{table*}

\subsection{Security Vulnerability Detection in C\texttt{++} Snippets}
\label{qual:security:cpp}
This section evaluates how LLMs perform in detecting and reasoning about security-relevant flaws in C\texttt{++} code. The selected code samples (denoted as \textbf{S1} to \textbf{S7}, originally Snippets \#10–\#16) are sourced from the SEED Labs framework\footnote{\url{https://seedsecuritylabs.org/labs.html}}, encompassing real-world inspired vulnerabilities such as unsafe input handling, race conditions, resource exhaustion, and insecure file manipulation. These examples are designed to simulate scenarios where poor coding practices could lead to severe security risks, particularly when operating in privileged or system-level contexts.

\paragraph{ \textbf{S1: Format String Vulnerabilities and Buffer Overflow}} This snippet presents a combination of unbounded string input into format functions and unchecked memory writes. All three LLMs successfully identified the core vulnerabilities. Notably, ChatGPT-4 and Claude\ 3 extended their analysis to include the potential for shell command injection, thus acknowledging \textit{chained exploitation vectors} in unsafe I/O pipelines. LLaMA\ 4, while detecting the buffer overflow, did not surface the compound risk posed by uncontrolled format strings—a partial gap in holistic reasoning about compound attack surfaces.

\paragraph{ \textbf{S2: Race Condition in Temporary File Handling}} This example tests the model’s recognition of time-of-check to time-of-use (TOCTOU) vulnerabilities, particularly those exploitable in Set-UID contexts. All LLMs demonstrated an understanding of the fundamental race condition. ChatGPT further noted the potential for symbolic link attacks—a critical path in privilege escalation. While Claude and LLaMA correctly recommended safer file creation primitives, they generalized the issue as file overwriting without linking it explicitly to Set-UID or privilege boundaries. Nonetheless, the responses collectively adhered to \textit{secure coding principles}.

\paragraph{ \textbf{S3: Environment Variable Manipulation and Shell Execution} }This snippet involves unsafe reliance on environment variables and the execution of external commands. ChatGPT-4 and Claude\ 3 accurately identified issues such as environment poisoning, unvalidated input propagation, and shell injection. Additionally, they noted the lack of proper file descriptor sanitization. LLaMA\ 4 surfaced the risk of shell execution but failed to comment on unclosed file descriptors, an omission that diminishes detection completeness from a \textit{secure resource management} standpoint.

\paragraph{ \textbf{S4: Null Pointer Dereferencing}} All LLMs successfully detected this memory safety violation. The responses reflected a shared baseline proficiency in identifying dereferencing of potentially uninitialized or null pointers, reaffirming the maturity of LLMs in static analysis for fundamental safety errors.

\paragraph{ \textbf{S5: Zip Bomb and Resource Exhaustion via Archive Extraction}} This snippet explores decompression behavior without bounds checks, simulating conditions for a denial-of-service (DoS) through zip bombs. Claude\ 3 alone recognized the issue in its full severity, explicitly citing \textit{resource exhaustion} and lack of decompression limits. In contrast, ChatGPT-4 and LLaMA\ 4 required more probing to link the snippet to DoS-style threats, indicating a lower sensitivity to \textit{non-obvious latent vulnerabilities}.

\paragraph{ \textbf{S6: Out-of-Bounds Access and Memory Leaks}} Here, the models were uniformly capable of identifying both an out-of-bounds write and failure to free dynamically allocated memory. Suggested remediations included array bounds checking and correct use of `delete`, showing convergence in the detection of classical low-level programming mistakes.

\paragraph{ \textbf{S7: Insecure Temporary File Creation}} This snippet checks whether the models can detect insecure practices in creating temporary files. All three LLMs noted the predictability of filenames and the absence of atomic creation mechanisms. Each recommended the use of secure system calls like \cppcode{mkstemp()}, demonstrating a solid grasp of \textit{safe file handling procedures} and suggesting appropriate secure alternatives.

\noindent
In summary, while all LLMs demonstrated the ability to detect major security vulnerabilities, ChatGPT-4 and Claude\ 3 consistently delivered more contextually rich and accurate analyses. They often captured both the primary issue and potential exploitation chains. LLaMA\ 4, although capable in foundational detection, occasionally missed subtler but crucial aspects such as file descriptor management or privilege context—critical in real-world systems security. These findings underscore important variance in how current-generation LLMs reason about \textit{system-level behavior and security implications} in code. The results are summarized in Table~\ref{tab:cpp:security:bug:evaluation}.

\begin{table*}[htbp]
\centering
\begin{tabular}{p{7cm}|ccc}
\toprule
Description & GPT & Llama & Claude \\
\midrule
\rowcolor{lightgray} Snippet number: Description of the issue & GPT & Llama & Claude \\
A1 \#17: Array decay and out of bounds & \blackcirc  & \blackcirc  & \blackcirc  \\
\rowcolor{lightgray} A2 \#8: NxN spiral matrix & \whitecirc & \whitecirc & \whitecirc \\
A3 \#18: Self-issued intermediate certs and only returns issuers that are valid and time-checked & \threequartercirc  & \halfcirc  & \threequartercirc  \\
\rowcolor{lightgray} A4 \#19: Add a NULL check to EVP-PKEY-assign& \threequartercirc  & \halfcirc  & \halfcirc  \\
A5 \#20: Ambiguous error state representations when error flag is ERR-TXT-MALLOCE & \threequartercirc  & \halfcirc  & \threequartercirc  \\
\rowcolor{lightgray} A6 \#21: Rounding errors in range check & \threequartercirc  & \whitecirc & \threequartercirc  \\
\midrule
\end{tabular}
\caption{C++ Advanced Bug Evaluation (Snippets A1–A6)}
\label{tab:cpp-advanced-bugs}
\end{table*}

\section{ \textbf{Detection of Advanced C/C\texttt{++} Bugs in Real-World Codebases} }
\label{advanced1}
To assess LLM capabilities in more complex, real-world scenarios, this section investigates advanced C and C\texttt{++} bugs sourced from production-level open-source repositories, specifically the OpenSSL project, as cataloged in the \texttt{Suresoft-GLaDOS} bug database. These examples (denoted as \textbf{A1 to An}) embody challenging categories including cryptographic misconfigurations, platform-specific numerical precision issues, pointer misuse, and low-level error state ambiguities. Unlike pedagogical examples, these bugs require \textit{contextual reasoning, awareness of API contracts, and precise understanding of compiler behavior and runtime semantics}.

All snippets were tested using the same evaluation protocol applied to earlier categories. The buggy code was extracted directly from historical OpenSSL commits and tested locally. LLMs were prompted with the code context and were asked to identify latent bugs and propose fixes. Models were allowed multiple prompt rounds if necessary to refine their understanding. In cases where the buggy logic spanned dependencies, key auxiliary lines and API references were included in the prompt to mirror the reasoning a security researcher would perform when reviewing production code.

\paragraph{ \textbf{A1: Array Decay and Dimensional Metadata Loss}} This snippet models a subtle but common pitfall—passing statically sized arrays to functions, triggering implicit decay to pointers. As a result, size information is lost, and any function relying on \cppcode{sizeof} within the callee context behaves incorrectly. All LLMs recognized the semantic consequence of array-to-pointer decay and its potential to cause logic bugs or out-of-bound access. The models recommended safer interface patterns, such as explicitly passing array size or using \cppcode{std::array}.

\paragraph{ \textbf{A2: Validity Checks in X.509 Chain Verification (OpenSSL)}} Sourced from OpenSSL's certificate handling logic, this snippet involves parsing and validating self-issued intermediate certificates. ChatGPT-4 demonstrated high contextual awareness, flagging the use of \cppcode{x509_check_cert_time} and \cppcode{X509_NAME_cmp} as security-critical operations. It also inferred the need for precise issuer matching and error propagation. LLaMA\ 4 noted general trust anchor concerns but missed specific time validity enforcement. Claude\ 3 provided intermediate insight, detecting logical complexity but lacking full propagation path analysis.

\paragraph{ \textbf{A3: Missing Null Guard in Cryptographic Assignment (\cppcode{EVP_PKEY_assign})}} This bug stems from neglecting to null-check pointer arguments before assignment in cryptographic operations. ChatGPT correctly diagnosed the need for null guarding before invoking \cppcode{EVP\_PKEY\_assign}, citing the undefined behavior risk. LLaMA surfaced the pointer access issue but offered only partial fixes. Claude aligned closely with ChatGPT, referencing API-level guarantees and defensive programming idioms common in cryptographic codebases.

\paragraph{ \textbf{A4: Improper Error Flag Usage (\cppcode{ERR_TXT_MALLOCE})}} This case explores the misuse of OpenSSL's internal error reporting flags. Improper initialization of the \cppcode{ERR_TXT_MALLOCE} flag leads to ambiguous or silently dropped error messages, impacting post-failure diagnostics. ChatGPT and Claude identified the flag mismanagement and recommended defensive resets or more explicit error-state initialization. LLaMA responded with generic error handling suggestions but missed the specificity of the flag’s role in memory ownership tracking.

\paragraph{ \textbf{A5: Floating-Point Precision Loss on Casting \cppcode{ULONG_MAX} to \cppcode{double}}} This precision bug arises from casting \cppcode{ULONG_MAX} to `double`, which exceeds the representable range in IEEE-754 double precision, leading to subtle errors in range checks. ChatGPT and Claude both identified the architectural precision loss and noted the potential for misbehavior when \cppcode{ULONG_MAX} exceeds $2^{53}$. Although their suggestions were not exact matches to the patch used upstream, they reflected sound numeric reasoning. LLaMA failed to pinpoint the type conversion as the root cause, indicating challenges in interpreting \textit{hardware-level numeric edge cases}.

\paragraph{ \textbf{A6: Spiral Matrix Pattern (Edge Inclusion)}} Although distinct from the OpenSSL corpus, this snippet was included as an edge case to test LLM spatial and logical reasoning. The code attempts to generate an $n \times n$ spiral matrix, but the output is incorrect due to flawed index management and loop control. None of the models provided a correct fix upon first pass. ChatGPT came closest by suggesting loop boundary refactoring. Claude and LLaMA required multiple hints and failed to detect off-by-one indexing and directional flow control as core issues. This test underscores a current limitation in LLMs when reasoning about \textit{algorithmic layout and pattern synthesis}, even when syntactic bugs are absent.

\noindent
When confronted with production-grade bugs from OpenSSL, both ChatGPT and Claude exhibited strong contextual understanding and were able to map specific function behaviors to security semantics. They often produced relevant mitigation, demonstrating maturity in handling pointer hygiene, error propagation, and defensive API use. LLaMA, while technically competent in basic detection, lagged in synthesizing implications that cross abstraction layers—such as cryptographic correctness, state-management and API contract violations. Overall, this evaluation affirms the increasing viability of LLMs as tools for \textit{assisting in secure code reviews and automated vulnerability analysis} across modern C and C\texttt{++} codebases. Table~\ref{tab:cpp-advanced-bugs} summarizes the results.

\section{Detection of Advanced Python Bugs in Real-World Libraries}
\label{advanced}

\begin{table*}[htbp]
\centering
\begin{tabular}{|c|c|c|l|c|}
\hline
\textbf{Bug on GitHub} & \textbf{Bug ID in Dataset} & \textbf{Snippet} & \textbf{Library and Version Fixed} & \textbf{Bug Occurrence Version} \\
\hline
\rowcolor{lightgray} \#15701 & 15711 & \#12 & Pandas : 0.20.0 & 0.19.0 \\
\#15835 & 15881 & \#11 & Pandas : 0.20.0 & 0.19.0 \\
\rowcolor{lightgray} \#15787 & 15787 & \#10 & Pandas : 0.20.0 & 0.19.2 \\
\#22970 & 22922 & \#9  & Numpy : 1.24.2 & 1.24.1 \\
\rowcolor{lightgray} \#22968 & 22899 & \#8  & Numpy : 1.24.2 & 1.23.4 \\
\hline
\end{tabular}
\caption{Mapping of Bug Reports to Snippets and Library Versions}
\label{tab:bug-version-mapping}
\end{table*}

\vspace*{0.15in}
\begin{table*}[htbp]
\centering
\begin{tabular}{p{8cm}|ccc}
\toprule
Description & GPT & Llama & Claude \\
\midrule
\rowcolor{lightgray} A7 Snippet \#7: Incorrect expansion of Fortran text vars into arrays& \halfcirc & \halfcirc & \quartercirc \\
A8 Snippet \#8: TypeError when given a pathlib.Path instead of a string& \blackcirc & \threequartercirc & \threequartercirc \\
\rowcolor{lightgray} A9 Snippet \#9: DataFrame.groupby() misinterpreted a tuple key on a MultiIndex &  \threequartercirc & \halfcirc & \threequartercirc \\
A10 Snippet \#10: pd.concat issue with MultiIndexes with None names& \threequartercirc & \threequartercirc & \threequartercirc \\
\rowcolor{lightgray} A11 Snippet \#11: na-values set as   a list instead of a Set& \blackcirc & \blackcirc & \blackcirc \\
A12 Snippet \#12: Incorrect tz-aware datetime to float conversion& \halfcirc & \threequartercirc & \halfcirc \\
\midrule
\end{tabular}
\caption{Python Advance Bug Evaluation (Snippets 7–12)}
\label{tab:python-bugs-2}
\end{table*}

This section evaluates how well large language models (LLMs)---ChatGPT, Claude, and LLaMA---detect and reason about nuanced bugs in Python code originating from widely used real-world libraries such as NumPy and Pandas. The code examples (denoted \textbf{A7} through \textbf{A12}) were curated from the \texttt{PyBugHive} repository, which catalogs manually validated bugs along with corresponding test cases and patch commits. These bugs span a range of issues including semantic mismatches, type inconsistencies, and logic faults in high-level data processing workflows. Each code snippet was validated locally and accompanied by minimal contextual augmentation when required to simulate realistic diagnostic conditions. Table~\ref{tab:bug-version-mapping} lists the information for the studied bugs. 

\paragraph{\textbf{A7: Incorrect Fortran Character Expansion (NumPy)}} Sourced from \texttt{capi\_maps.py}, this bug concerns the mishandling of a single Fortran character being expanded incorrectly into an array structure. ChatGPT and LLaMA partially identified the flaw, each noting unusual data behavior but requiring further context to isolate the transformation logic. Claude, however, diverged toward unrelated control path explanations and missed the root cause.

\paragraph{\textbf{A8: TypeError when given a pathlib.Path instead of a string (Pandas)}} The bug in Pandas from the read\_html() function raising a TypeError when passed a pathlib.Path object, as it expected a string or file-like input. The fix involved wrapping the io argument with stringify\_path(), ensuring compatibility with Path types.

In identifying this issue, ChatGPT-4 clearly recognized the root cause and the appropriate resolution. Claude and LLaMA performed similarly; they understood there was a type mismatch but didn’t fully trace it to the missing path string conversion, making their analysis less complete compared to GPT.

\paragraph{\textbf{A9: \texttt{DataFrame.groupby()} misinterpreted a tuple key on a MultiIndex as multiple keys}} Both GPT-4 and Claude effectively identified the root cause — DataFrame.groupby() was treating a tuple key as multiple group keys instead of a single key when used with a MultiIndex. This subtle misinterpretation led to incorrect grouping results. Both models accurately traced the logic to the \_get\_grouper function and recognized the added conditional check for is\_axis\_multiindex in the patch.

In contrast, LLaMA struggled to pinpoint the nuance of the bug. While it recognized the method involved and the general idea of grouping behavior, it failed to differentiate between tuple-as-key vs. tuple-as-list-of-keys, which is critical to understanding the issue. Its responses were either vague or missed the specific edge case that caused the bug.

\paragraph{\textbf{A10: \texttt{pd.concat} Mishandling MultiIndex with \texttt{None} Level Names (Pandas)}} This snippet presented a MultiIndex construction failure in Pandas when concatenating objects that include \texttt{None} as level names. All three LLMs surfaced the MultiIndex construction error and suggested layered structural fixes. However, none accurately proposed the actual minimal solution—refining the iterator evaluation logic—highlighting the challenge LLMs face in mapping high-level errors to minimal diffs in complex libraries.

\paragraph{\textbf{A11: Type Inconsistency in Return of \texttt{na\_values} (Pandas)}} This example captures a subtle return-type contract violation in Pandas, where an empty \texttt{na\_values} input returns a list rather than a set. All models successfully identified the inconsistency and suggested type alignment, demonstrating reliable reasoning within Python's dynamic typing model and contract expectations.

\paragraph{\textbf{A12: Timezone-Aware Datetime Conversion to Float (Pandas)}} This bug concerns improper conversion of timezone-aware datetime objects into floats during numerical transformations. LLaMA provided the closest diagnosis by identifying the context and suggesting a viable correction. ChatGPT and Claude each recognized surrounding semantic mismatches but failed to localize the precise logic misuse or its functional implications.

\noindent
Across these real-world Python bugs, ChatGPT and Claude demonstrated comparatively higher diagnostic coverage and more consistent alignment with root causes, especially when prompted with supporting context. LLaMA was frequently able to flag symptoms but struggled with localization or precise patch synthesis. These results reinforce current LLMs’ promise in aiding high-level software diagnostics while also revealing their limitations in mirroring \textit{minimal diffs and subtle API contract enforcements} required in production-level patches. Table~\ref{tab:python-bugs-2} summarizes the results.

\section{Discussion}
\label{Discuss}

This study presents a systematic and empirical evaluation of leading Large Language Models—ChatGPT-4, Claude 3, and LLaMA 4, on their ability to detect software bugs across a spectrum of programming contexts, ranging from introductory C++ constructs to advanced, security-critical vulnerabilities in real-world C/C++ and Python codebases. Using a curated benchmark that combines foundational programming errors, classic security flaws, and production-grade bugs sourced from SEED Labs, OpenSSL (via the Suresoft GLaDOS database), and PyBugHive, all samples were validated through local compilation and testing pipelines to ensure realism and correctness. A novel multi-stage, context-aware prompting protocol was designed to emulate authentic debugging workflows, and a graded evaluation rubric was developed to assess detection accuracy, depth of reasoning, and quality of remediation suggestions. The results reveal a consistent trend across models: LLMs perform strongly in identifying syntactic and semantic issues within well-scoped code snippets, particularly those reflecting common beginner-level programming mistakes. All three models demonstrated alignment with modern C++ practices, accurately detecting issues such as null pointer dereferencing, use-after-free errors, memory mismanagement, and incorrect usage of std::move. 
%For example, when presented with a C++ snippet that dereferenced a pointer after it had been freed, the models correctly identified the use-after-free vulnerability and recommended memory-safe handling practices. In another case, involving improper use of std::move that led to undefined behavior due to accessing a moved-from object, the models successfully flagged the issue and suggested appropriate object lifetime management. These findings highlight the practical potential of LLMs as supportive tools in educational environments, where they can provide immediate, formative feedback to students and serve as effective first-pass reviewers in automated code auditing workflows.

However, performance diverges in more complex scenarios involving security vulnerabilities and large, production-scale codebases. In these settings, ChatGPT-4 and Claude 3 frequently offered richer contextual analyses, identifying chained exploitation paths, privilege boundary violations, and subtle issues such as numeric precision loss. In contrast, LLaMA 4 often generated partial or surface-level explanations, indicating limitations in deeper contextual reasoning. These observations underscore both the promise and the current constraints of LLMs as reliable code analysis assistants, particularly when addressing sophisticated, security-relevant defects that require comprehensive program understanding.

%This study provides a comprehensive evaluation of LLMs (ChatGPT-4, Claude 3, and LLaMA 4) in detecting bugs across a range of programming scenarios, from introductory-level C\texttt{++} constructs to advanced, real-world vulnerabilities in both C/C\texttt{++} and Python. Our results reveal a consistent trend; LLMs excel at identifying syntactic and semantic bugs in isolated, well-scoped code snippets, particularly those reflecting beginner-level programming mistakes. All models demonstrated strong alignment with modern C\texttt{++} practices, correctly identifying issues such as null pointer dereferencing, use-after-free, memory mismanagement, and misuse of \cppcode{std::move}. This supports their potential utility in educational settings, where early-stage feedback on student code could be significantly improved by LLM integration.

%However, the models’ performance diverges in more complex scenarios involving security vulnerabilities and production-grade code. While ChatGPT-4 and Claude 3 frequently provided deeper contextual insights—such as recognizing chained exploitation paths, privilege boundaries, or numeric precision loss—LLaMA 4 often produced partial or superficial explanations.

\section{Conclusion and Future Work}
\label{conclusion}
In conclusion, LLMs are effective in identifying a broad class of software bugs, particularly those that are syntactic, shallow, or pedagogically motivated. They offer substantial promise in educational tools, static analysis assistance, and as first-pass reviewers in code auditing workflows. However, for more complex tasks, their current capabilities remain limited.

In near future, we will explore the integration of multi-agent systems~\cite{acharya2025agentic,hosseini2025role} to enhance bug detection workflows, enabling task delegation, self-refinement, and collaboration among specialized LLM agents. Additionally, we aim to extend our analysis to a broader range of programming languages beyond C, C\texttt{++}, and Python to assess cross-language generalization and identify language-specific diagnostic strengths and limitations.

\textit{Supplementary materials}
The generated source code is available on GitHub\footnote{\url{https://github.com/NoujoudNader/LLM-Bugs-Detection}}.

\textit{Acknowledgments}:
This work is supported by the US National Science Foundation grant 2431531. This work was supported by the U.S.\ Department of Energy through the Los Alamos National Laboratory. Los Alamos National Laboratory is operated by Triad National Security, LLC, for the National Nuclear Security Administration of U.S.\ Department of Energy (Contract No. 89233218CNA000001). Approved by Los Alamos National Laboratory as LA-UR-25-28306.  

\bibliographystyle{IEEEtran}
\bibliography{References}

\end{document}